\documentclass[twocolumn,superscriptaddress,reprint,amsmath,amssymb,aps,prb]{revtex4-1}
\usepackage{graphicx}
\usepackage{color}
\usepackage{dcolumn}
\usepackage{hyperref}
\usepackage{epstopdf}
\usepackage{amsmath}
\hfuzz=\maxdimen
\usepackage{float}

\begin{document}

\title{Z$_2$ nontrivial topology of rare-earth binary oxide superconductor}

\author{Jiahui Qian}
\affiliation{State Key Laboratory of Surface Physics and Department of Physics, Fudan University, Shanghai 200433, China}

\author{Zongqi Shen}
\affiliation{State Key Laboratory of Surface Physics and Department of Physics, Fudan University, Shanghai 200433, China}

\author{Xinyuan Wei}
\affiliation{State Key Laboratory of Surface Physics and Department of Physics, Fudan University, Shanghai 200433, China}

\author{Wei Li}
\email{w$\_$li@fudan.edu.cn}
\affiliation{State Key Laboratory of Surface Physics and Department of Physics, Fudan University, Shanghai 200433, China}
\affiliation{Collaborative Innovation Center of Advanced Microstructures, Nanjing University, Jiangsu 210093, China}

\date{\today}

\begin{abstract}
Recently, superconductivity has been discovered in rock-salt structured binary lanthanum monoxide LaO through state-of-the-art oxide thin-film epitaxy. In this work, we reveal that the normal state of superconducting LaO is a $Z_2$ nontrivial topological metal, where the Dirac point protected by the crystal symmetry is located around the Fermi energy. By analysing the orbital characteristics, we show that the nature of the topological band structure of LaO originates from the intra-atomic transition from the outer shell La 5$d$ to the inner shell 4$f$ orbitals driven by the strong octahedral crystal-field. Furthermore, the appearance of novel surface states unambiguously demonstrates the topological signature of LaO superconductor. Our theoretical findings not only shed new light into the understanding of the exotic quantum behaviors in LaO superconductor with intimate correlation between 4$f$ and 5$d$ orbitals in La, but also provide an exciting platform to explore the interplay between nontrivial topology and superconductivity.
\end{abstract}

\maketitle

{\it Introduction.---}The search for exotic quantum states in quantum materials is an appealing subject in condensed matter physics. Among them, rare-earth compounds show unique physical properties such as mixed-valence phenomena~\cite{CMVarma1976,Lawrence1981}, Kondo insulating states~\cite{Mason1992,Dzero2010}, heavy-fermion behavior~\cite{Stewart1984}, and intriguing superconductivity~\cite{Mathur1998}. In particular, the rare-earth binary compound of lanthanum superhydride LaH$_{10}$ under extreme high pressures displays high transition temperature ($T_c$) superconductivity with a $T_c$ of 260 K, which represents a step toward the goal of achieving room-temperature superconductors~\cite{Somayazulu2019, Drozdov2019}. At ambient pressure, the binary lanthanum dicarbide LaC$_2$ shows superconductivity at 1.6 K~\cite{Green1969}. Compared with LaC$_2$, the binary lanthanum sesquicarbide La$_2$C$_3$ is characterized by a higher $T_c$ of 11 K~\cite{Giorgi1969,Francavilla1976}, which can be further enhanced to 13 K by tuning the La/C ratio~\cite{Kim2006}. The inherent pairing symmetry of these binary lanthanum carbide superconductors, however, needs further scrutiny~\cite{Smidman2017}.

Recently, a new type of binary lanthanum monoxide LaO with rock-salt structure was reported to be a superconductor with a $T_c$ of 5 K using state-of-the-art oxide thin-film epitaxy techniques~\cite{Kaminaga2018}, since rare-earth elements usually prefer their trivalent ionic state to form the most stable binary sesquioxides~\cite{Mikami2006}, which are well-known high permittivity wide-gap insulators for gate dielectrics~\cite{Wilk2001}. For metastable rare earth monoxides, the electronic properties are significantly different from those of sesquioxides due to the presence of outer shell 5$d$ conduction carriers in La atoms, which strongly interact with their localized inner shell 4$f$ electrons around the Fermi energy~\cite{Adachi1998}. Electronic band structure calculations have shown that the conducting electrons around the Fermi level in LaO superconductor mainly come from the contribution of La 5$d$ orbitals, strongly hybridized with the La 4$f$ orbitals~\cite{PHSun2021}. This implies that the binary lanthanum monoxide superconductor LaO may lead to topologically nontrivial electronic structures driven by the La 5$d$-4$f$ orbital transitions with opposite parity.

In this paper, we revisit and discuss the nature of the interplay between La 5$d$ and 4$f$ orbitals in the newly discovered rare-earth binary superconductor LaO using the first-principles calculations. Our results show that the normal state of LaO superconductor is a nontrivial $Z_2$ topological metal. Interestingly, the topologically protected Dirac point is located around the Fermi energy. This originates from the inherent features of the LaO, i.e. from the La 5$d$-4$f$ orbital transitions driven by the strong octahedral crystal-field. The existence of novel surface states further unambiguously demonstrates the topological signature of LaO superconductor in its normal state.

{\it Theoretical Calculations.---}In our first-principles calculations~\cite{Liwei2012,Liwei2014,Liwei2020}, we employ the plane-wave basis method and the Perdew-Burke-Ernzerhof exchange correlation potential~\cite{PBE}, as implemented in the VASP code~\cite{VASP}. A 600 eV cutoff in the plane wave expansion and a $20\times 20\times 20$ Monkhorst-Pack $\mathbf{k}$-grid are chosen to ensure accuracy up to $10^{-5}$ meV. The lattice constants are optimised to make forces on individual atoms below $10^{-3}$ eV/\AA. Furthermore, we take into account the contribution of spin-orbit coupling using the second variational method~\cite{Yuan2021}, and 
the phonon dispersion is calculated using the density functional perturbation theory within the phonopy code~\cite{Togo2015,Huang2018}.

\begin{figure}
\centering
\includegraphics[bb=120 30 445 450,width=5.5cm,height=6.8cm]{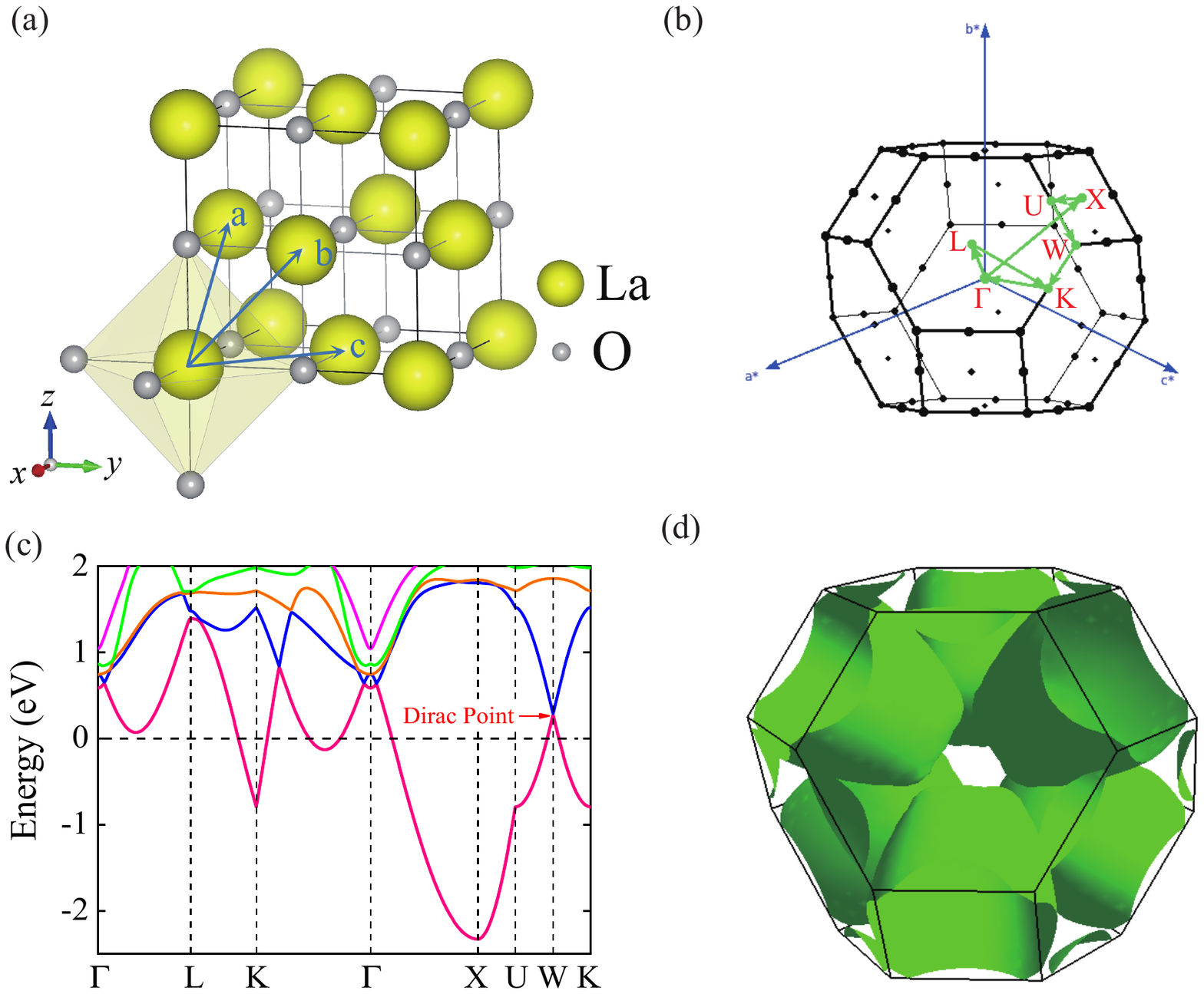}
\caption{(a) Crystal structure of LaO. Here \textbf{a}, \textbf{b}, and \textbf{c} are the primitive lattice vectors. (b) BZ for the primitive cell of LaO. The high-symmetry paths are indicated by the green arrows. (c) Electronic band structure of LaO along the high-symmetry paths shown in (b). The Fermi energy is set to zero in (c). (d) Corresponding Fermi surface topology.
}\label{fig1}
\end{figure}

Fig.~\ref{fig1}(a) shows the crystal structure of LaO superconductor, which crystallizes in a face-centered-cubic (fcc) structure of rock-salt type with space group \textit{Fm-3m}. The optimized lattice constants of LaO are $5.158$ \AA, in good agreement with previous experimental values ($5.144$ \AA)~\cite{Leger1981}, and with the theoretical prediction ($5.164$ \AA) obtained using an ultrasoft pseudopotential~\cite{PHSun2021}. The corresponding Brillouin zone (BZ) of the primitive cell of LaO along the high-symmetry $\mathbf{k}$-point is depicted in Fig.~\ref{fig1}(b). The stability of the optimized lattice structure of LaO is also examined by calculating the phonon spectrum, shown in Fig. S1~\cite{SI}. 
There are no imaginary frequencies in the phonon spectrum, confirming the dynamical stability of the rock-salt structured lanthanum monoxide LaO.

The calculated electronic band structure of LaO is shown in Fig.~\ref{fig1}(c) (also see Fig. S2~\cite{SI}). There is a single-band passing through the Fermi level $E_f$ across the BZ, and this suggests a metallic behavior of LaO superconductor in the normal state [also see the Fermi surface topology in Fig.~\ref{fig1}(d)]. This result is intuitively expected since for the La$^{2+}$ ion in the monoxide LaO, the nominal number of the configuration La$^{2+}$: 5$d^1$4$f^0$ electron is 1, and thus the lowest-lying conduction band is partially occupied. In the vicinity of the Fermi level $E_f$, there is an intriguing Dirac point (0.2 eV above the $E_f$ about extra 0.2 $e^-$) appearing at the high symmetric $W$ point in BZ~\cite{Wehling2014,Young2012}. Usually, the appearance of a Dirac point in three-dimensional systems (preserving both the time-reversal and inversion symmetries), where two Weyl points overlap in momentum space, is unstable. On the other hand, the crystal symmetry of $C_{4v}$ at $W$ point in BZ protects the stability of Dirac point in LaO (see the strained band in Fig. S3)~\cite{SI,LFu2011,Song2017}. In addition, we also notice that there is a band crossing above $E_f$ (at 0.56 eV), indicating the existence of band inversion in the electronic band structure of LaO. Overall, these results strongly indicate that the electronic structure of LaO superconductor in the normal state is a nontrivial topological metallic phase in three dimensions~\cite{Wehling2014}.

\begin{table}[t!]
\caption{Parities $\delta^{\kappa}_{\Gamma_i}$ at the $\kappa$-th band and the eight time-reversal invariant $\mathbf{k}$ points [one $\Gamma$ $(0, 0, 0)$,  three $X$ $(\frac{1}{2}, \frac{1}{2}, 0)2\pi/a$, and four $L$ $(\frac{1}{2}, \frac{1}{2}, \frac{1}{2})2\pi/a$] for LaO. Due to the Kramers degeneracy, we only list the parities on the even numbers bands. The nontrivial $Z_2$ topological invariants are evaluated through the product of the parities $\delta^{\kappa}_{\Gamma_i}$ at the eight time-reversal invariant $\mathbf{k}$ points on a selected band with index of $\kappa$ corresponding to that shown in Fig.~\ref{fig1}(c). 
}\label{tableParity}
\begin{ruledtabular}%
\begin{tabular}{ccccc}
 & $\delta_{\Gamma}$ & $3\times \delta_X$  & $4\times \delta_L$ & $Z_2$ \tabularnewline
\hline
\textcolor{green}{Band 24}    & -1  & -1  & +1  & (0;000) \tabularnewline
\textcolor[RGB]{255,165,0}{Band 22}   & +1  & -1  & -1  & (1;000) \tabularnewline
\textcolor{blue}{Band 20}     & +1  & -1  & -1  & (1;000) \tabularnewline
\textcolor{magenta}{Band 18}  & +1  & +1  & -1  & (0;000) \tabularnewline
\end{tabular}\end{ruledtabular}
\end{table}

Since LaO is invariant under time-reversal symmetry (see the magnetization calculations listed in Table S1, with magnetic configurations shown in Fig. S4)~\cite{SI,Kane2005,LFu2007,Moore2007,Roy2009}, we have carried out the calculations of the $Z_2$ topological index and quantitatively inspected the nontrivial topological phases of LaO. Besides the spatial inversion symmetry in rock-salt structured LaO, the calculation of the $Z_2$ topological invariant can be dramatically simplified by the so-called ``parity method''~\cite{TI2007,Ortiz2020,Ortiz2021}. Accordingly, the $Z_2$ topological invariant of LaO can be obtained from the wavefunction parities $\delta^{\kappa}_{\Gamma_i}$ on a band index $\kappa$ around the Fermi level $E_f$ at the eight time-reversal invariant $\mathbf{k}$ points defined as $\Gamma_{i=(n_1n_2n_3)}=\frac{1}{2}(n_1\mathbf{a}^*+n_2\mathbf{b}^*+n_3\mathbf{c}^*)$. Here $\mathbf{a}^*$, $\mathbf{b}^*$, and $\mathbf{c}^*$ are the primitive reciprocal lattice vectors shown in Fig.~\ref{fig1}(b) with $n_j$=0 or 1, and determined by the quantities $\delta^{\kappa=2\kappa'}_{\Gamma_i}
=\langle\Phi_{2\kappa',\Gamma_i}|\hat{P}|\Phi_{2\kappa',\Gamma_i}\rangle=\pm 1$. In turn, they are the eigenvalues of the parity operator $\hat{P}$, and correspond to even (odd) parity of the Bloch functions $|\Phi_{2\kappa',\Gamma_i}\rangle$ at the $2\kappa'$-th band and time-reversal invariant $\mathbf{k}$ point $\Gamma_i$. 
Notice that we use the band numbering scheme from the calculations of density functional theory because LaO is a metal with no clear division between conduction and valence states, as shown in Fig.~\ref{fig1}(c)~\cite{Ortiz2020,Ortiz2021,Kasinathan2015}. For the LaO fcc lattice, we have the following eight time-reversal invariant $\mathbf{k}$ points: one $\Gamma$ $(0, 0, 0)$,  three $X$ $(\frac{1}{2}, \frac{1}{2}, 0)2\pi/a$, and four $L$ $(\frac{1}{2}, \frac{1}{2}, \frac{1}{2})2\pi/a$, shown in Fig.~\ref{fig1}(b). The calculated parities on LaO at these time-reversal invariant $\mathbf{k}$ points with a given band index $\kappa(=2\kappa')$ around the Fermi level $E_f$ are shown in Fig.~\ref{fig1}(c) (Kramers degenerate bands, $\delta_{\Gamma_i}^{2\kappa'}=\delta_{\Gamma_i}^{2\kappa'-1}$) and listed in Table~\ref{tableParity}. The $Z_2$ topological invariants $(\nu_0;\nu_1\nu_2\nu_3)$ between each pair of bands ($\kappa$) near the Fermi level are thus evaluated as the product of these parities at the time-reversal invariant $\mathbf{k}$ points. They are defined as $(-1)^{\nu_0}=\Pi_{i=1}^{8}\delta_{\Gamma_i}^{\kappa}$ and $(-1)^{\nu_k}=\Pi_{n_k=1,n_{j\neq k}=0,1}\delta_{\Gamma_{i=(n_1n_2n_3)}}^{\kappa}$, where $\nu_0$ is a strong topological index independent of the choice of primitive reciprocal lattice vectors $\Gamma_j$. On the contrary, $\nu_1$, $\nu_2$, and $\nu_3$, refereed to as weak topological indexes, do depend on the choice. Interestingly, the existence of nonzero topological indices (see Table~\ref{tableParity}), allows us to identify the normal state of superconducting LaO as a $Z_2$ topological metal (also see the detailed band topological invariants in Fig. S5 and Table S2)~\cite{SI}. The intriguing Dirac point and band inversion shown in Fig.~\ref{fig1}(c) occur at the transition between the trivial and nontrivial strong topological bands (see Band indexes 18 and 20 listed in Table~\ref{tableParity}), which confirms the inherent topological nature of the electronic state behaviors in LaO superconductor. 




\begin{figure}
\centering
\includegraphics[bb=65 480 450 790,width=8cm,height=6.5cm]{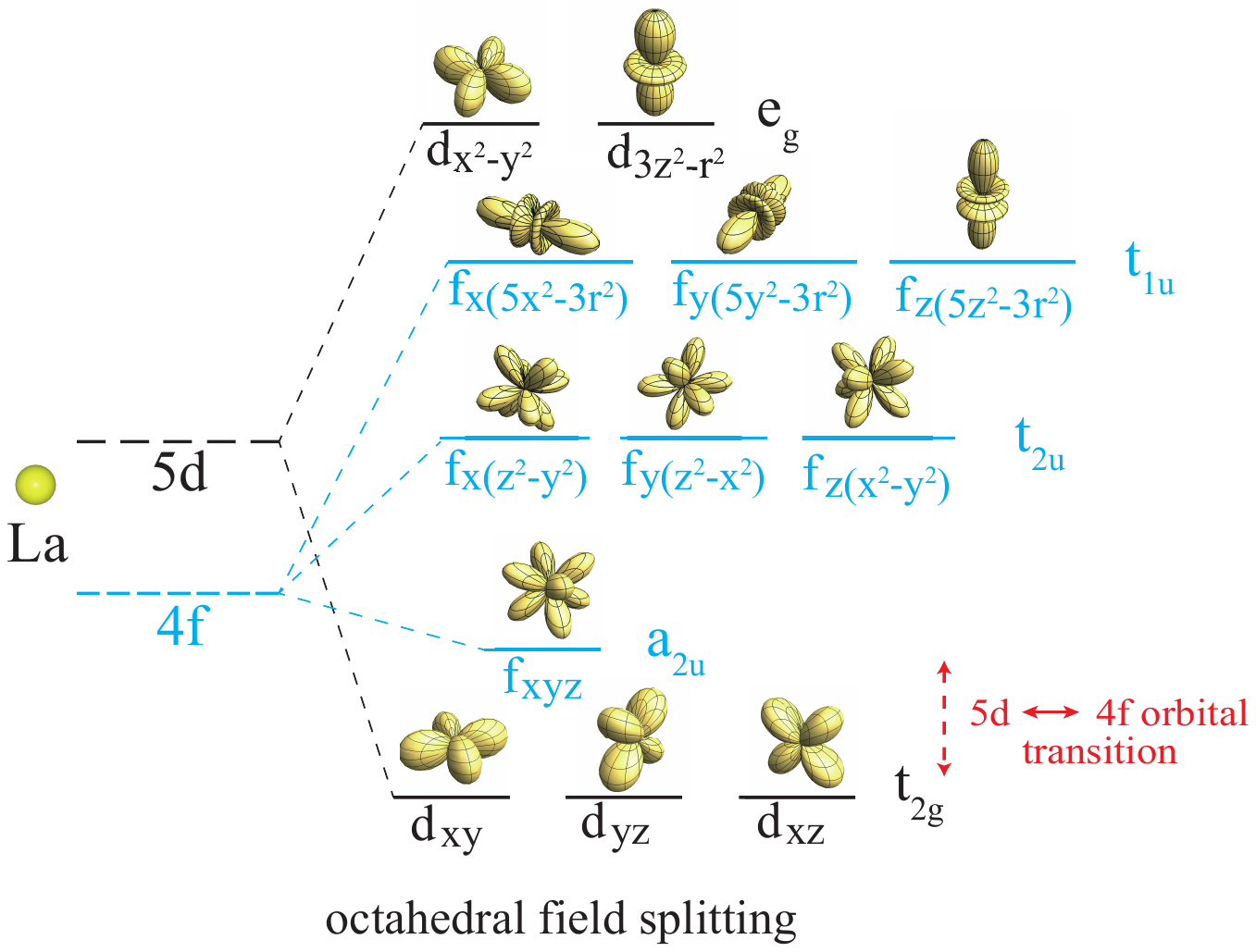}
\caption{Schematic diagram of the orbital evolution from the atomic outer shell 5$d$ to the inner shell 4$f$ orbitals of La driven by the octahedral crystal-field splitting in LaO with rock-salt structure at the $\Gamma$ point in BZ. The lobes of corresponding orbitals under the symmetry of octahedral field are also sketched. 
}\label{fig2}
\end{figure}

Having established the presence of a nontrivial topology in LaO, we now proceed to elaborate on the nature of this $Z_2$ topological metallic phase. From the viewpoint of ligand field theory~\cite{GroupTheory,Liwei2012}, the lanthanum in rock-salt structure LaO is coordinated by the oxygen octahedron [see Fig.~\ref{fig1}(a)]. The crystal field would normally split the outer shell five degenerate 5$d$ orbitals in La into the lower-lying threefold degenerate $t_{2g}$ composed of $d_{xy}$, $d_{yz}$, and $d_{xz}$ orbitals, and the upper-lying twofold degenerate $e_{g}$ composed of $d_{x^2-y^2}$ and $d_{(3z^2-r^2)}$ orbitals, as schematically illustrated in Fig.~\ref{fig2}. Furthermore, the inner shell seven degenerate 4$f$ orbitals in La are also be split simultaneously by the octahedral crystal-field into two groups of triply-degenerate orbitals [$t_{1u}$ composed of $f_{x(5x^2-3r^2)}$, $f_{y(5y^2-3r^2)}$, and $f_{z(5z^2-3r^2)}$ orbitals, and $t_{2u}$ composed of $f_{x(z^2-y^2)}$, $f_{y(z^2-x^2)}$, and $f_{z(x^2-y^2)}$ orbitals] and a lower-lying one nondegenerate orbital ($a_{2u}: f_{xyz}$ orbital) (also see Fig.~\ref{fig2}). However, by considering that the charge distribution of lower-lying orbitals $t_{2g}$ do not directly point to the oxygen anions 
in octahedral crystal-field, and hence their Coulomb interaction is relatively weaker than that for $a_{2u}$ orbital, we have that the energy of $t_{2g}$ levels is below that of $a_{2u}$. As a result, the system undergoes an orbital transition between $t_{2g}$ of 5$d$ and $a_{2u}$ of 4$f$ orbitals, accompanied by parity-switching (see Fig.~\ref{fig2}). For the La$^{2+}$ ion, the nominal number of conducting electrons is 1, which partially occupies the lowest-lying conducting band. Notice that the spin-orbit coupling has been neglected in the discussions on the orbital splitting shown in Fig.~\ref{fig2} because LaO is a single-band metal shown in Fig.~\ref{fig1}(c) and the spin-orbit coupling will be significantly quenched. We have also examined the electronic band structure without inclusion of spin-orbit coupling and found that the electronic dispersion near the Fermi level remains qualitatively unaltered (see Fig. S6)~\cite{SI}.
This orbital splitting picture is analogue to that used in interpreting topological Kondo insulating state in SmB$_6$ and YbB$_{12}$~\cite{Dzero2010,FLu2013,Dzero2016}. 
Overall, we conclude that the strong octahedral crystal-field splitting drives the intra-atomic transition in energy from $5d$ to 4$f$ orbitals of La, making the normal state of LaO superconductor a nontrivial $Z_2$ topological metal.



\begin{figure}
\centering
\includegraphics[width=9.2cm]{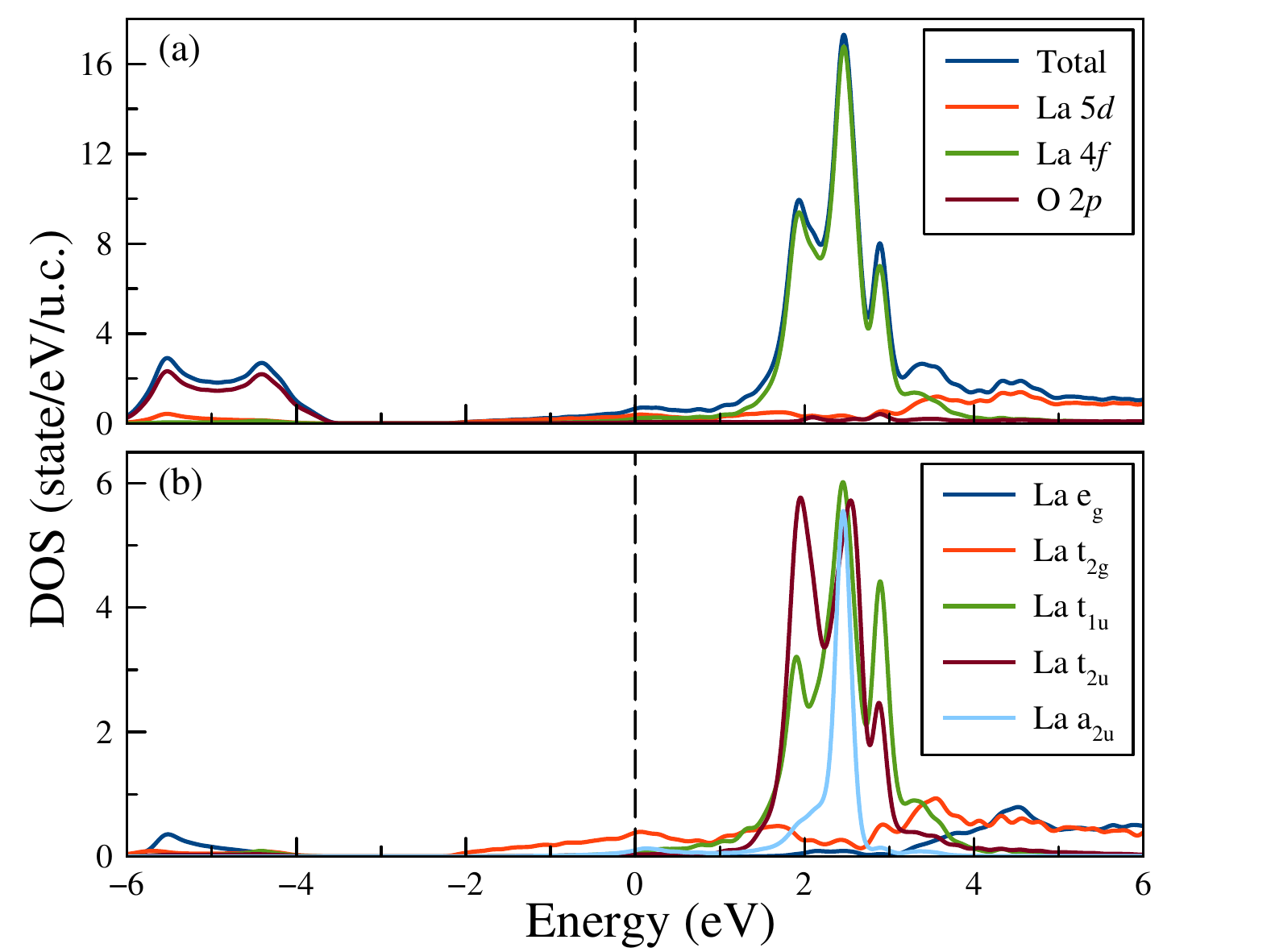}
\caption{(a) Total and projected DOS of La 5$d$ and 4$f$ and O 2$p$ orbitals for LaO. (b) Projected DOS of La 5$d$ and 4$f$ orbitals in the presence of the octahedral field. The Fermi energy is set to zero.
}\label{fig3}
\end{figure}

The density of states (DOS) are also evaluated to gain more insight into the aforementioned orbital splitting picture shown in Fig.~\ref{fig2}. In Fig.~\ref{fig3}(a), we find that the electronic states around the Fermi level $E_f$ are mainly due to the contribution of La 5$d$ orbitals hybridized strongly with La 4$f$ orbitals. This result is consistent with the nature of electron configuration of 5$d^1$4$f^0$ of the La$^{2+}$ ion in the monoxide LaO, and is responsible for the intra-atomic transition from outer shell $5d$ to the inner shell 4$f$ orbitals of La. Additionally, it is important to point out that there is a van Hove singularity in the vicinity of the Fermi level $E_f$ with a DOS of 0.7 states per eV per La atom, which induces an effective reduction of the local electron Coulomb interaction, thereby enhancing the role of non-local electron correlations for the driving force of the appearance of superconductivity~\cite{Fleck1997}. Upon inspecting the orbital resolved DOS on La 5$d$ and 4$f$ orbitals, shown in Fig.~\ref{fig3}(b), we find that the La 4$f$ orbitals mainly locate inside the octahedral splitting gap $\Delta_0\approx 2.5$ eV between the upper-lying $e_g$ and lower-lying $t_{2g}$ orbitals split from the La 5$d$ orbitals. This is consistent with the orbital splitting picture schematically illustrated in Fig.~\ref{fig2} and with the fact that the octahedral crystal-field splitting gap for 5$d$ orbitals is significantly larger than 4$f$ ones. In the vicinity of the Fermi level $E_f$, we also notice that the $t_{2g}$ orbitals play a dominant role, followed by the $a_{2u}$ orbital. This corresponds to the trivial and nontrivial strong topological bands listed in Table~\ref{tableParity}, based on the orbital resolved electronic structure calculations, indicative of the $a_{2u}$ orbital across the lower-lying $t_{2g}$ orbitals resulting in the appearance of topologically protected Dirac point (also see the orbital resolved band in Fig. S7)~\cite{SI}.

\begin{figure}
\centering
\includegraphics[width=8cm,height=5.5cm]{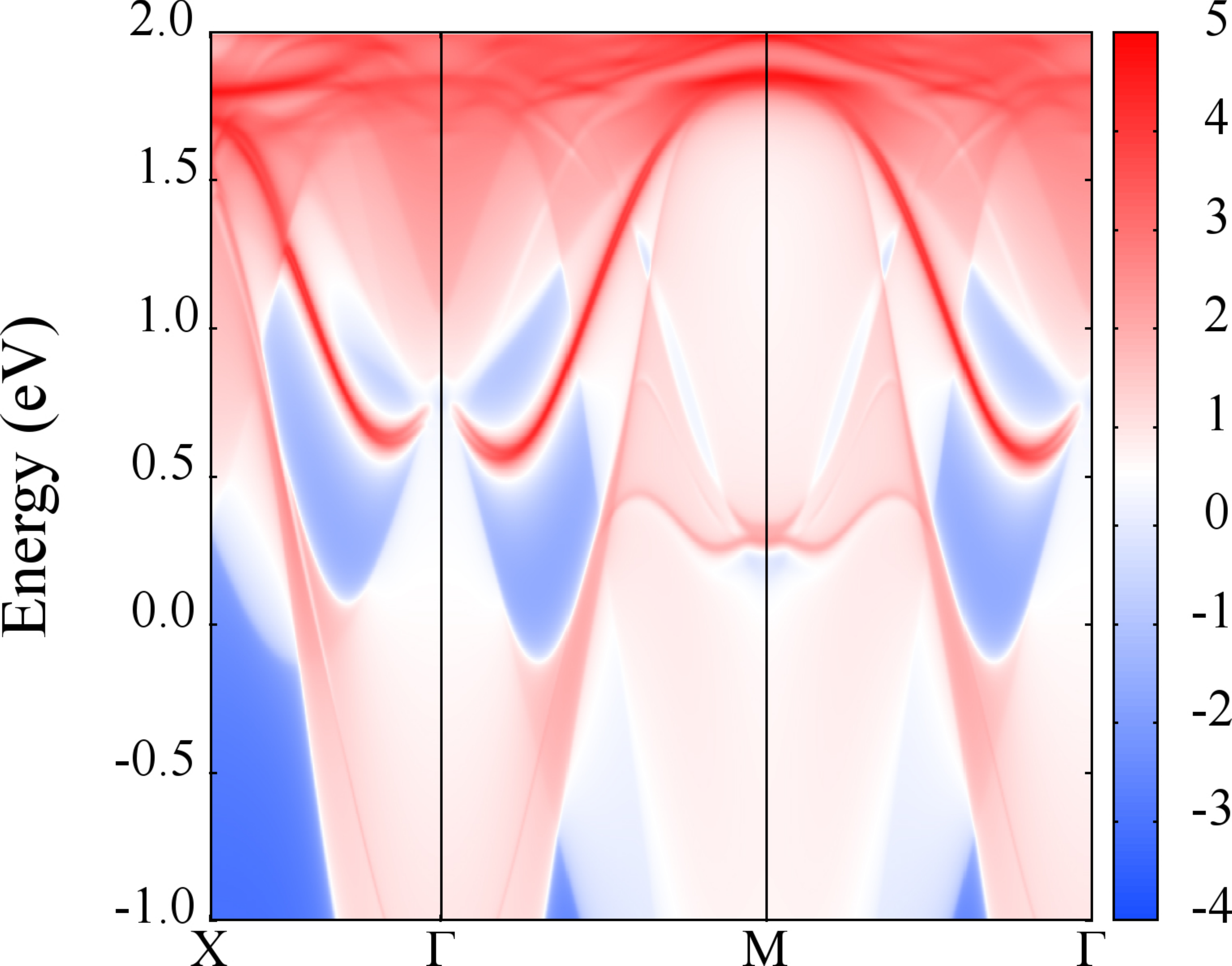}
\caption{The surface states of LaO projected on the $(001)$ plane. Warmer colors denote higher local DOS and blue regions indicate the bulk band gap. The Fermi energy is set to zero.
}\label{fig4}
\end{figure}

At last, we discuss the topological surface states exploiting the idea of bulk-edge correspondence of the $Z_2$ topological metal~\cite{Halperin,Hatsugai}. In particular, we calculate the topological surface states of a semi-infinite system using the surface Green's function~\cite{Sancho1985,Liwei2011} constructed from the maximally localized Wannier function and fitting the first-principles band structure calculations~\cite{Marzari,Souza}. The dispersion of the surface states projected on the $(001)$ plane, shown in Fig.~\ref{fig4}, is obtained by evaluating the imaginary part of the surface Green's function~\cite{Liwei2014,QWu2018}. Upon comparison with the bulk band structure along $[001]$ direction (shown in Fig. S8)~\cite{SI}, one clearly sees the Dirac-like topological states at $M$ point, slightly above the Fermi level $E_f$ shown in Fig.~\ref{fig4}. By unfolding the BZ, one sees that this is actually projected from the $W$ point shown in Fig.~\ref{fig1}(c). Additional surface states may be seen around the center of BZ ($\Gamma$ point), indicated by bright spots, which stem from the inverted band shown in Fig.~\ref{fig1}(c) and associated with nonzero $Z_2$ number listed in Table~\ref{tableParity}. These results about surface states further reveal the nontrivial topological electronic band structure of superconducting LaO in its normal state, and can be experimentally detected by angle-resolved photoemission spectroscopy (ARPES).

{\it Discussions.---}The nontrivial $Z_2$ topological band structure of three-dimensional superconducting LaO is of great interest for stabilizing the formation of Majorana modes within the vortex cores of a natively superconducting surface state that are relevant for fault-tolerant quantum computing~\cite{LFu2008,Nayak2008,Alicea,Beenakker2013}. Materials hosting both topologically nontrivial surface states and a native superconducting ground state are uncommon, with relatively few promising candidates identified in layered FeSe$_{1-x}$Te$_x$~\cite{GXu2016,SZhu2019} and recently discovered layered Kagome superconductor CsV$_3$Sb$_5$~\cite{Ortiz2020,Ortiz2021}. Additionally, it is worth to point out that although the topologically protected Dirac point is located slightly above the Fermi level, the LaO system can be easily driven into such region with relatively light electron doping [about 0.2 $e^-$ shown in Fig.~\ref{fig1}(c)]. This may be achieved by changing the amount of oxygen vacancies in LaO thin films during deposition, which may be also used to effectively tune the critical temperature $T_c$~\cite{Kaminaga2018}. On the other hand, since in three-dimensional systems the Dirac point around Fermi level $E_f$, see Fig.~\ref{fig1}(c), is usually unstable, a stable and topological protected Weyl points may originate from Dirac point by breaking either time-reversal or inversion symmetries~\cite{Young2012,Wehling2014}. We thus suggest that Weyl superconductors may be realized using LaO superconductor capped by EuO ferromagentic insulator using the proximity effect of ferromagnetism to lift the time-reversal symmetry. Alternatively, one may exploit isovalent substitution of sulfur to break the inversion symmetry~\cite{TMeng2017}, which can be experimentally verified by $in$ $situ$ ARPES in the near future.

{\it Conclusion.---}Using the first-principles calculations on the newly discovered lanthanum monoxide superconductor (LaO), we have found that the strong electron correlation between 4$f$ and 5$d$ orbitals in La leads to the emergence of nontrivial topological band structure in its normal state, such as topologically protected Dirac point around the Fermi energy and the appearance of topological signature of novel surface states. Our theoretical study provides crucial insights into the electronic structure of superconducting LaO, hereby laying down the basis for a solid understanding of the interplay between nontrivial topology and superconductivity.

{\it Acknowledgments.---}This work was supported by the National Natural Science Foundation of China (Grant Nos. 11927807 and 61971143) and Shanghai Science and Technology Committee (Grant Nos. 18JC1420400 and 20DZ1100604). J.Q. acknowledged the computational resources at Fudan University and Suzhou University of Science and Technology.


\end{document}